# Liquid Hole Multipliers: bubble-assisted electroluminescence in liquid xenon

L. Arazi[1,*], E. Erdal[1,†], A. E. C. Coimbra[1,2], M. L. Rappaport[1], D. Vartsky[1], V. Chepel[2] and A. Breskin[1]

  [1]*Department of Particle Physics and Astrophysics*
  *Weizmann Institute of Science, Rehovot 7610001, Israel*
  [2] *Department of Physics, University of Coimbra, 3004-516 Coimbra, Portugal*
  E-mail: lior.arazi@weizmann.ac.il

ABSTRACT: In this work we discuss the mechanism behind the large electroluminescence signals observed at relatively low electric fields in the holes of a Thick Gas Electron Multiplier (THGEM) electrode immersed in liquid xenon. We present strong evidence that the scintillation light is generated in xenon bubbles trapped below the THGEM holes. The process is shown to be remarkably stable over months of operation, providing – under specific thermodynamic conditions – energy resolution similar to that of present dual-phase liquid xenon experiments. The observed mechanism may serve as the basis for the development of Liquid Hole Multipliers (LHMs), capable of producing local charge-induced electroluminescence signals in large-volume single-phase noble-liquid detectors for dark matter and neutrino physics experiments.

---

[*] Corresponding author
[†] Equal contributor

# Contents

# 1   Introduction

Dual-phase noble liquid time projection chambers (TPCs) [1-3] are presently leading the effort for direct detection of dark matter in the form of weakly interacting massive particles (WIMPs). They contain a large volume of liquid target (liquid xenon, LXe, or liquid argon, LAr) in equilibrium with its vapor phase. Radiation interacting with the liquid leads to a prompt scintillation signal (S1). Ionization electrons liberated at the site of interaction drift in an electric field towards the liquid surface where they are extracted into the gas phase and produce a delayed electroluminescence signal (S2) proportional to the number of extracted electrons. Both S1 and S2 are usually recorded by top and bottom arrays of vacuum photomultiplier tubes (PMTs).

Ongoing dual-phase experiments include XENON100 [4, 5], LUX [6, 7] and PandaX-I [8, 9] (all using LXe), as well as DarkSide-50 [10] (using LAr). While current dark matter detectors utilize up to a few hundred kg of LXe or LAr as their sensitive targets and probe the spin-independent WIMP-nucleon cross section down to $10^{-45} – 10^{-46}$ cm$^2$, ton-scale dual-phase experiments are already under construction or in different planning stages, e.g., XENON1T [11] (later to be upgraded to XENONnT), ArDM [12], LZ [13] and DarkSide G2 [10]; these are expected be sensitive to cross sections down to $10^{-47} – 10^{-48}$ cm$^2$. Multi-ton detectors, such as DARWIN [14], are foreseen to supersede these experiments in the next decade, reaching 10-fold higher WIMP detection sensitivities, to the point where solar and atmospheric neutrinos will become the dominant background through coherent neutrino-nucleus scattering.

A primary challenge in scaling dual-phase detectors to the multi-ton regime is the requirement to keep the liquid surface parallel to and between two mesh electrodes that should, themselves, be completely parallel to one another across a few meters. This is needed in order that the ratio of S2 to the extracted charge will be constant over the area of the electrodes [2], which is essential for the detector's background discrimination capability (especially in LXe-based detectors). S2 scales linearly with the product of the number of electrons extracted into the gas phase, the width of the gas gap and the electric field across the gap [3]; therefore, tilt, surface waves, variations in the liquid level and local deformations in the mesh electrodes can seriously increase the spread in S2 for constant extracted charge and, hence, reduce the capability to discriminate between candidate WIMP-like events and background-induced events.

A possible solution to this problem is to move from a dual-phase TPC configuration to a *single-phase* liquid-only TPC scheme in which S2 signals would be generated by electrodes immersed within the liquid, rather than in the vapor phase [15, 16]. One implementation, suggested in [16] and studied in [17] and [18], relies on thin (Ø5-20 μm) anode wires. In this scheme ionization electrons released at the point of interaction drift towards the immersed wires, where they induce secondary scintillation (accompanied by modest charge multiplication) in the intense field close to the wire surface. Recent measurements with Ø5 μm and Ø10 μm wires [17] demonstrated an S2 scintillation yield of ~300 photons per drifting electron and a charge gain of ~10-15, in agreement with previous works from the 1970s [19-22].

A second approach, first suggested in [23], is to use immersed hole-multipliers, such as GEMs [24] or THGEMs [25, 26], to generate S2 signals inside the liquid. It was suggested that in such electrodes, termed "Liquid Hole-Multipliers" (LHMs), ionization electrons would be focused into the holes where the field may be high enough to induce proportional scintillation; this, in turn, would be detected by photosensors (e.g., PMTs or gaseous photomultipliers – GPMs [27]) located behind the LHM. Similarly, coating the LHM electrode with cesium iodide could permit the detection of S1 photons by the production of photoelectrons in the photocathode; these would be subsequently focused into the holes to produce secondary scintillation in the same process. It was further suggested that LHMs may be deployed in cascaded structures, where the signal would be propagated between stages by electroluminescence photons, resulting in increased photon yields and possibly allowing for charge readout [23].

When the LHM idea was first brought up, it was thought that even though the electric field inside the electrode's holes would be considerably lower than that close to the surface of thin wires, the much longer trajectory of the drifting electrons through it might still result in an appreciable scintillation yield. Indeed, preliminary experiments with an alpha particle source and a single THGEM electrode immersed in LXe [28] suggested an S2 yield of ~600 photons per drifting electron, in a modest field of ~30 kV/cm at the hole center (an estimate based on the assumption that the major contribution to the electroluminescence signal originates from the hole's center). This field value was more than 10-fold lower than the threshold field for scintillation reported for thin wires [17, 21, 29]; it was, however, of the same order of magnitude as similar observations with a THGEM [30] and GEM [31] electrodes immersed in LAr, where proportional scintillation was observed at a field of ~50-60 kV/cm.

Further experiments, on which we report here, substantiated the observation of proportional scintillation from a THGEM electrode immersed in LXe. However, the response of the S2 signals to changes in the thermodynamic conditions in the system (in particular to abrupt pressure changes), led us to conclude that the light signals are in fact generated by electroluminescence in a *xenon gas layer or bubbles* trapped below the holes of the THGEM, rather than in the liquid phase. The electroluminescence process was shown to be surprisingly stable over months of operation, with unexpectedly good energy resolution for certain combinations of pressure and temperature and reproducible results. In what follows we describe in detail the evidence leading to the bubble hypothesis, discuss the steady state and transient behavior of the electroluminescence process, and study its dependence on the applied fields. We conclude by suggesting possible schemes for the potential application of the *bubble-assisted LHM* in future noble-liquid TPCs.

## 2  Experimental setup

The experiments were conducted using the Weizmann Institute Liquid Xenon (WILiX) system, described in detail in [28] and shown schematically in figure 1. The system comprises an inner LXe chamber (inner diameter 346 mm) within an outer vacuum chamber. Xenon condenses inside the LXe chamber on the fins of a temperature-controlled Cu cold finger, cooled by a Brooks Automation PCC cryocooler. A 50 W cartridge heater located at the lower part of the cold finger is regulated by a Cryo-con Model 24C controller to keep the fin temperature constant to within ±1 mK. Continuous xenon recirculation is provided by a double-diaphragm pump (ADI DiaVac Pump, model R061, with EPDM diaphragms) through a SAES MonoTorr hot getter, model PS4-MT3-R2. Xenon gas returns to the inner chamber through a commercial parallel-plate heat exchanger (GEA model GBS100M), where it is cooled and condensed at ~95% efficiency by the extracted liquid. Most of the volume of the inner chamber is taken up by a cylindrical PTFE block, instrumented by Pt100 temperature sensors and capacitive liquid-level gauges. The experiments were conducted with the cold finger fins set to temperatures in the range 163-176 K, with the chamber pressure ranging, accordingly, from 1.3 to 2.3 bar. The Xe gas flow was typically 3 slpm.

The central region of the PTFE block housed the LHM setup, shown in figure 2. The LHM electrode in this work was the same as in [28]: a circular THGEM electrode with an active diameter of 34 mm, consisting of a 0.4 mm thick FR4 substrate with a hexagonal array of Ø0.3 mm drilled holes with 0.1 mm etched rims and with a 1 mm pitch. The 20 μm Cu clad on the two faces of the THGEM electrode was Au plated. The THGEM electrode was held in place by PTFE spacers. A non-spectroscopic 16.4 kBq $^{241}$Am source with an active diameter of 4 mm was held above the THGEM. The distance between the active surface of the source and the THGEM top face was 3.3±0.1 mm (taking into account the shrinkage of PTFE at LXe temperatures). The source was fixed to a stainless steel disc with high-voltage bias, ~30 mm below the liquid surface. A ring of holes around the source allowed for LXe circulation through the LHM assembly. The voltage difference between the source holder and the top face of the THGEM electrode defined the drift field, $E_{drift}$. An electro-formed Cu mesh with 85% transparency (Precision Eforming, MC17), consisting of a square grid of 18.5 μm wires with a pitch of 340 μm was installed 2.5 mm below the THGEM electrode; the voltage difference between the THGEM bottom face and the mesh defined the nominal transfer field, $E_{trans}$. All four surfaces – source holder, THGEM top, THGEM bottom and mesh – were biased separately using a CAEN N1471H HV power supply unit. Light signals were recorded by a Hamamatsu

PMT, model R8520-06-Al-MOD (QE of 22.4% at 175 nm, as quoted by the manufacturer for room temperature), located 6.5 mm below the mesh. For the major part of the study the PMT was biased at 600 V (negative bias on the photocathode, anode at ground), ensuring operation in its linear regime over the range of observed S2 signals. The PMT signals were fed directly (without amplification) into a digital oscilloscope (Tektronix 5054B); recorded waveform files were processed off-line with dedicated software tools developed for this purpose. The typical sampling rate was 250 MS/s, with an acquisition rate of ~1000 frames per second.

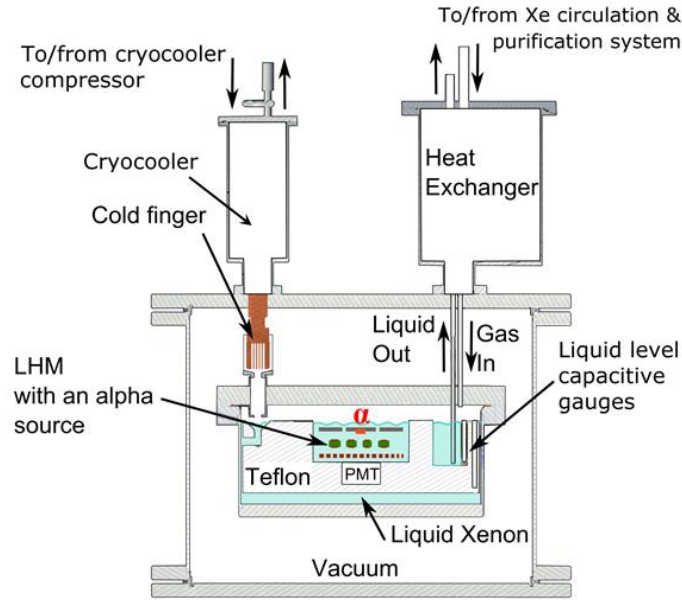

**Figure 1**: Schematic drawing of the WILiX liquid xenon cryostat, with the LHM assembly (not to scale) at its center.

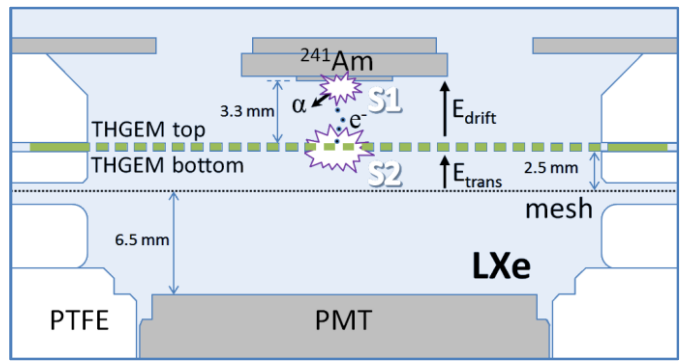

**Figure 2**: Schematic drawing of the central part of the LHM setup. The main components are drawn to scale. Alpha particle emission into the liquid results in prompt scintillation light (S1). Ionization electrons liberated along the alpha particle track drift towards the THGEM holes where they induce secondary scintillation (S2). The 175 nm light from both S1 and S2 is detected by the PMT. The liquid level (not shown) is ~30 mm above the $^{241}$Am source.

## 3   Results

### 3.1   Establishing S2 in steady-state

A prerequisite for observing S2 signals is a sufficiently high LXe purity, to avoid electron capture along their drift to the THGEM. With freshly filled LXe, before starting the purification process, only S1 signals were observed. S2 signals appeared within three days after starting LXe recirculation and purification at a flow of 2.6 slpm, with the system in steady state at 1.3 bar. The signals were observed on the digital oscilloscope at a constant rate with an S2 signal accompanying ~99.5% of the S1 triggers (the trigger level was set well above the PMT dark counts). The average magnitude of S2 signals reached its steady state within nine days after starting LXe recirculation and remained steady over the following four months. Given the short drift time from the source to the THGEM, it is safe to conclude that the loss of ionization electrons to electronegative impurities was negligible throughout the experiments.

Figure 3 shows a typical waveform with S1 and S2 signals; in this particular case the THGEM voltage was 1500 V, the drift field was 0.5 kV/cm and the nominal transfer field was 1 kV/cm (pulling the electrons towards the bottom mesh). The drift time depended on the drift field and pressure, and was found to be in good agreement with calculations based on drift velocity data from the literature [32]. The typical S2 signal rise-time, reflecting the electroluminescence process, was ~150-250 ns (depending on the drift field and the THGEM voltages), with a typical FWHM of ~300 ns.

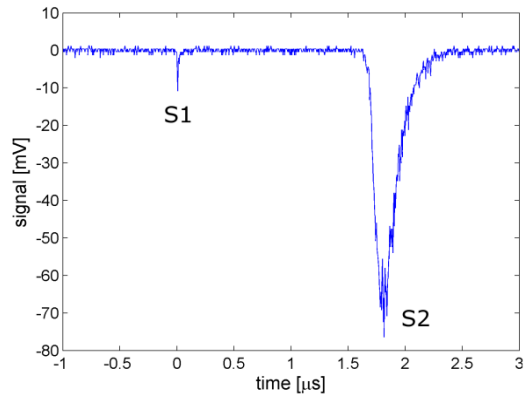

**Figure 3**: A typical sample waveform of alpha particle-induced S1 and S2 recorded by the PMT (setup of figure 2).

### 3.2   S2 response to pressure and temperature changes

Having reestablished the constant presence of S2 signals as in [28] and suspecting that their origin was scintillation in xenon gas below the THGEM electrode, we proceeded to study their response to changes in the system pressure and temperature. The expectation was that a rapid increase in the system pressure would result in the disappearance of S2 due to bubble collapse, and that the signals would reappear once bubble formation resumes.

Changes in the cryostat pressure and temperature can be induced by modifying the temperature of the cold finger condensing the xenon vapor (figure 1). A step increase of the cold finger set point leads to an immediate reduction in the rate of vapor condensation and an increase in pressure. The temperature of the inner parts of the cryostat follows such changes slowly (because of their large heat capacity), with a typical time scale of several hours. To study how this affects the S2 signals, we recorded their rate in conjunction with the main thermodynamic parameters of the system. For the rate measurement, the PMT signals, shaped by a timing-filter amplifier and processed by a discriminator, were fed into the counter input of an NI-DAQ USB60008 card; the discriminator threshold was set to detect solely S2 signals (S1 signals remained below threshold).

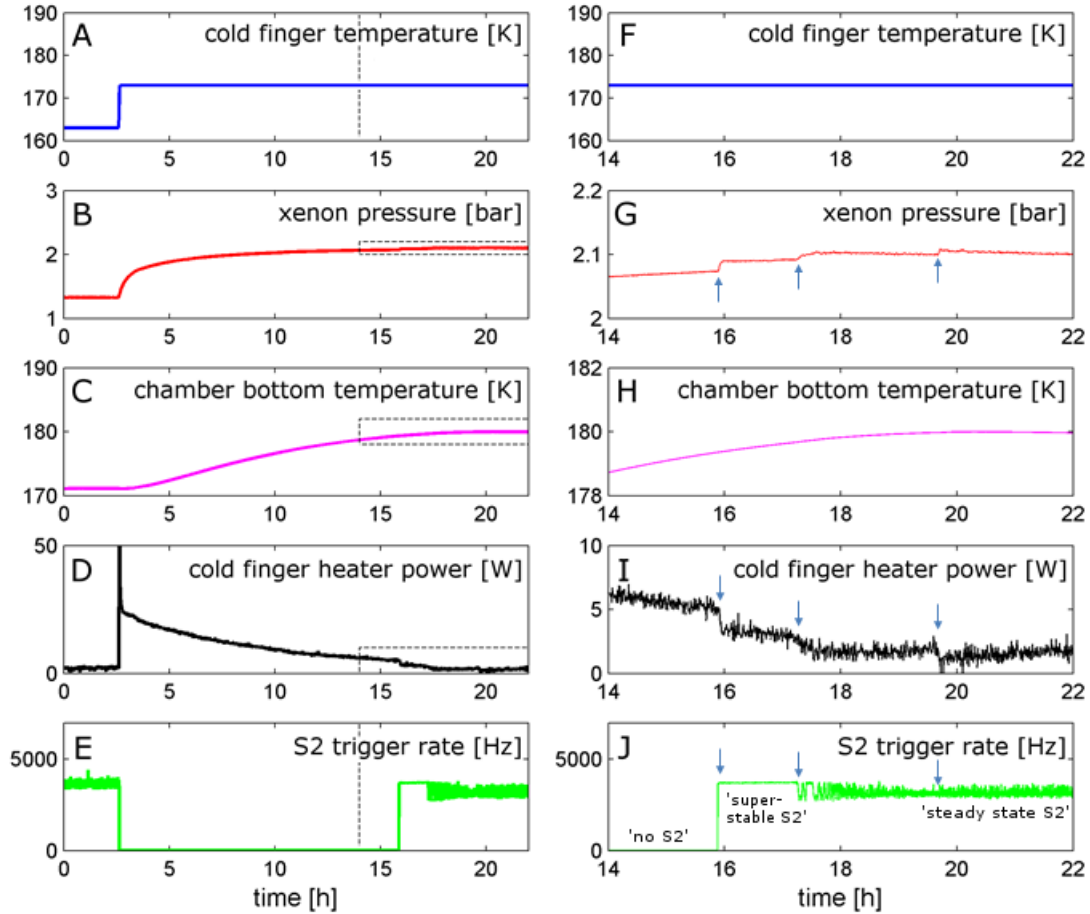

**Figure 4**: Main thermodynamic parameters of the LXe cryostat and S2 trigger rate as a function of time following a step increase in the cold finger temperature set point. (A) Cold finger temperature; (B) xenon pressure; (C) temperature at the LXe chamber bottom; (D) power provided by the cold finger heater (to keep the fin temperature at 173 K); (E) S2 trigger rate. (F)-(J) show expanded views of (A)-(E), focusing on the last 8 hours of the experiment (and corresponding to the dashed rectangles on the left). Arrows in (G), (I) and (J) mark correlated changes in pressure, heater power and S2 trigger rate. Note the 'super-stable' period from t=15.9 h to t=17.3 h in (J).

Figure 4 shows the behavior of the system over ~20 hours following a step increase of the cold finger set point from 163 K to 173 K. The pressure starts rising immediately after increasing the set point. This is accompanied by a rapid drop of the S2 trigger rate to zero (no S2 signals are observed on the oscilloscope at this stage). Over the following 20 hours, the temperature at the chamber's bottom gradually increases from 172 K to 180 K. About 13 hours after increasing the cold finger temperature, S2 signals reappear.

The behavior of the S2 signals is consistent with the bubble hypothesis − during the initial steady state (before increasing the cold finger set point) there is a constant presence of bubbles below the THGEM, leading to constant S2 signals. When the cold finger temperature increases, the rapid rise in pressure causes the existing bubbles to collapse; the formation of new bubbles is inhibited, because the vapor pressure of xenon corresponding to the temperature of the inner parts is lower than the pressure above the liquid. With no bubbles under the THGEM, there are no S2 signals. Once the temperature of the bubble-forming surface(s) becomes high enough, bubble formation resumes, new bubbles accumulate below the THGEM, and the S2 signals reappear. Note that in WILiX, in order to allow bubbles to grow against the hydrostatic pressure (and surface tension) at a depth of a few cm, the temperature of the liquid at the site of bubble formation should be higher by only a fraction of a degree than that at the surface [33]. Indeed, at the moment S2 reappeared (t = 15.9 h), the temperature measured at the chamber bottom was higher by only 0.8 K than that corresponding to the measured gas pressure.

During the initial steady state, the rate of S2 triggers fluctuates (E). These fluctuations are attributed to instabilities in the S2 waveform pulse height, also seen on the oscilloscope (see additional details in section 3.3.2 below). Surprisingly, once the S2 signals reappear (at t=15.9 h in (J)), there is a regime lasting ~1.4 hours (in this particular experiment) where the S2 rate is very stable; following that, it returns to the usual steady state fluctuation regime (at t=17.3 h). Interestingly, the transitions from the 'no S2' regime to this 'super-stable S2' regime and then back to the 'steady state S2' regime, are coincident with small step increases in the gas pressure and small downward steps in the power supplied by the heater to maintain the cold fins at 173 K, as indicated by the arrows in (G), (I) and (J) (a third downward step in the heater power accompanies an apparent 'waist' in the S2 rate at t=19.7 h). This may indicate that the transitions in the S2 pattern reflect sequential changes in the bubble formation process (perhaps occurring on several surfaces), where each change is accompanied by an increase in the total heat transfer from the bubble-forming surface(s) into the liquid.

Over several months of measurements, in thermodynamic steady state (constant pressure and temperature), the system was found to be *always* in the 'steady state S2' mode at pressures spanning the range 1.3-2.3 bar (the system could not be kept at higher pressures in steady state due to cooling power limitations, and lower pressures were avoided to prevent freezing of the xenon). In this regime, S2 signals were observed for nearly all S1 triggers (>99.5%); the rare "S1 with missing S2" events (at a rate of a few Hz) may be attributed to energy depositions occurring below the THGEM by shallow-angle cosmic muons or radioactive emissions from the system components. Transitions to the 'no S2' condition and back to 'steady state S2' were repeatedly demonstrated by performing step increases and decreases in the cold finger set point. During the transient 'no S2' condition, S2 signals were completely absent, even for THGEM voltages as high as ~4 kV, where occasional discharges occurred. In almost every case, the reappearance of steady state S2 was preceded by a 'super-stable' period (following a few-hour-long period of 'no S2'); however, on one occasion the transition to the 'super-stable' mode required a second increase of the cold finger temperature by 1 K.

### 3.3  Study of S1 and S2 properties

The confirmation of our bubble-assisted electroluminescence hypothesis in LXe was followed by a systematic study of this new, unexpected, operation mechanism. The study focused mainly on the parameters affecting the scintillation yield within the THGEM holes and the S2 pulse resolution; this included investigating the roles played by the drift field, THGEM voltage, transfer field, and xenon pressure.

### 3.3.1 S1 spectrum, event selection and S2 pulse-height resolution

The $^{241}$Am alpha source used in this experiment included a ~2 μm thick protective layer; as a result, the most probable value of the emitted alpha particle energy spectrum was reduced from 5.5 MeV to 4.4 MeV and the spectrum was asymmetric, with a prominent lower-energy tail. Other emissions from the source included 59.5 keV gammas (emitted in coincidence with alpha particles in 36% of the decays), as well as additional gammas and conversion electrons with lower energies.

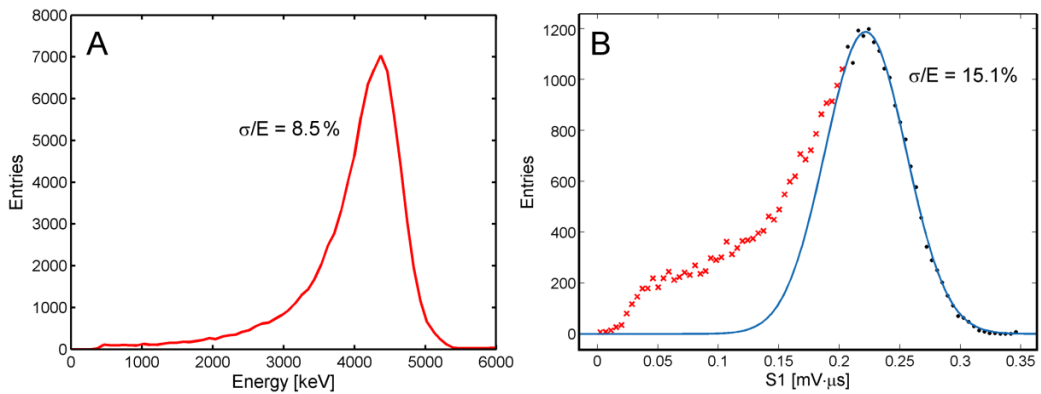

**Figure 5**: The spectrum of alpha particles emitted from the non-spectroscopic $^{241}$Am source. (A) Measurement with a Si surface barrier detector, including alpha particles emitted with angles up to ~45° to the source normal. (B) Measured S1 spectrum of alpha particles emitted into LXe, using the PMT below the THGEM (see figure 2); the RMS resolution is for a Gaussian fitted to the right side of the peak. The tail extending to lower energies results from partial energy deposition inside the source protective layer. The knee on the left in (B) results from the finite trigger threshold.

Figure 5A shows a measurement of the source spectrum using a Si surface-barrier detector recording direct alpha particle hits; the setup geometry in this case limited the data to alpha particles emitted with angles of up to ~45° from the normal. Figure 5B shows a scintillation-based measurement of the alpha particle spectrum in the LHM setup shown in figure 2; here, the PMT detected S1 light from the tracks of the alpha particles emitted over $2\pi$ steradians into the liquid. An alpha particle stopped in LXe spends, on average, 18 eV per VUV photon [3]; taking into account light transmission through the THGEM electrode holes, we estimate that the emission of a 4.4 MeV alpha resulted in ~500 photons impinging on the PMT (this value is later used as a reference to estimate the S2 yield in section 3.3.5). Considering the PMT's QE and photoelectron collection efficiency, we estimate that 4.4 MeV alpha emissions into the liquid resulted in S1 signals of ~60-70 photoelectrons; the emission of 59.5 keV gammas typically did not result in a detectable S1 signals (with less than one photoelectron on average).

Figure 6 shows a 2D histogram of S1 and S2 waveform areas (time integral of the pulses) recorded in the 'super-stable S2' condition (see figure 4J) with a drift field of 0.5 kV/cm, 1.25 kV across the THGEM and a nominal transfer field of 1 kV/cm, at 2.1 bar; the PMT voltage was 600 V. Triggers were either on alpha particle S1, alpha particle S2, or S2 signals associated with 59.5 keV gammas or low-energy gammas and electrons (for gammas and electrons, S1 was at the level of electronic noise). The histogram shows the main features of the $^{241}$Am source discussed above. The central diagonal distribution is associated with the alpha particles (of which the majority are in coincidence with gammas and/or conversion electrons). The 59.5 keV and the low-energy gamma and conversion electron "blobs" are clearly seen. Although differing by a factor of ~70 in energy, the S2 ratio between alphas and the 59.5 keV gammas is only ~4. This is attributed to the large difference in the charge yield (number of ionization electrons escaping recombination per keV) between an alpha particle track and a track of a recoiling electron in LXe [34, 35].

The study presented next focuses on alpha particle-induced events. As can be inferred from figure 6, the selection of such events can be done easily by setting a threshold on the S1 signal area. In what follows, we chose S1 > 0.1 mV·μs, which allowed for good statistics at the price of including some events with lower energies, resulting in a somewhat asymmetric S2 spectrum. Additional cuts were employed to discard events including more than one S2 peak, or events in which the S2 peak appeared to comprise two overlapping contributions; this allowed rejecting a substantial number of events in which the emission of an alpha particle into the liquid was accompanied by a coincident emission of a 59.5 keV gamma. After applying the above cuts, the centroid and variance of the S2 distribution were extracted from a Gaussian fit to the right side of the peak, to suppress the residual asymmetry of the spectrum.

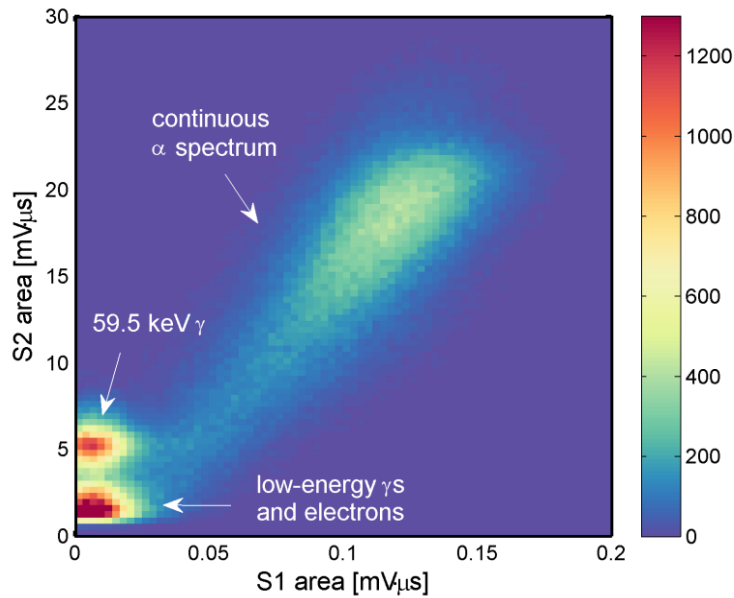

**Figure 6:** S2 vs. S1 2D histogram recorded under the 'super-stable S2' conditions at 2.1 bar with 0.5 kV/cm drift field, 1.25 kV across the THGEM and a nominal transfer field of 1 kV/cm. The histogram shows the main emission channels of the non-spectroscopic $^{241}$Am source.

Figure 7 demonstrates the event selection methodology by showing the S2 spectrum (here acquired under 'super-stable S2' conditions) with no S1 cut, with a cut requiring S1 > 0.1 mV·μs, and with additional alpha-gamma coincidence cuts. The small shoulder on the right side of the alpha peak is attributed, at least in part, to events with coincident alpha and 59.5 keV gamma emissions. The full spectrum comprises 416,395 waveforms, acquired during ~1 hour of super-stable conditions (same dataset as in figure 6). A Gaussian fit to the right side of the peak (after applying the coincidence cuts) resulted in σ/E = 11.1%. For comparison, the measured S2 resolution of the XENON100 experiment for a similar number of ionization electrons (there, for 236 keV gammas), is 10.0 ± 1.5 % [4].

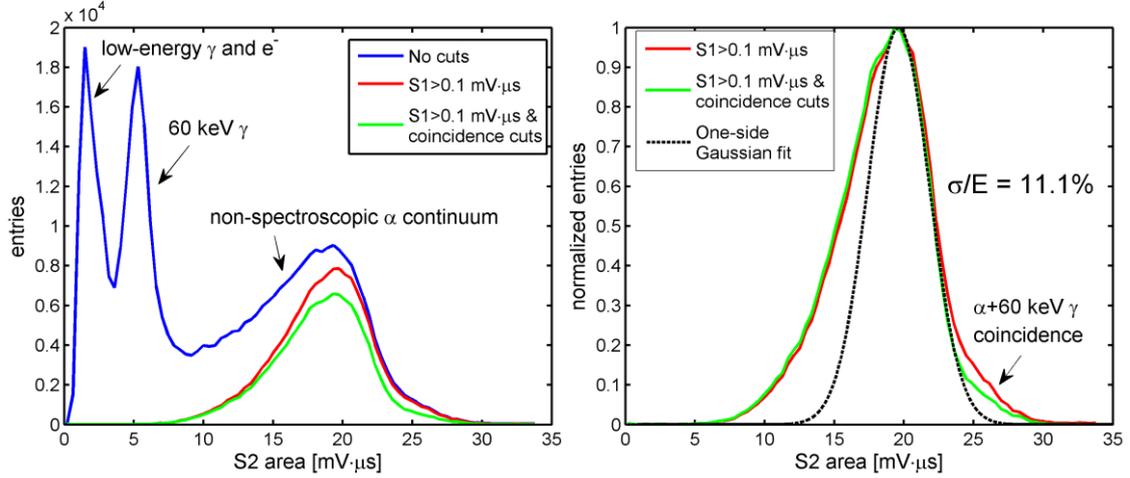

**Figure 7:** Measured S2 spectra for different applied cuts (left) for 'super-stable S2' conditions. A high threshold on S1 area removes the lower-energy gamma and electron peaks, along with most of the low-energy tail of alpha emissions. Right: the alpha-particle S2 spectra normalized to their maximum, for the S1 cut and gamma coincidence cut, with a right-side Gaussian fit.

*3.3.2    S2 properties under 'steady state S2' conditions*

Unlike the 'super-stable S2' state, considerable fluctuations in S2 magnitude were observed in steady state (as evident in figure 4). The fluctuations appeared to come in occasional high-amplitude bursts, each lasting between a few seconds to a few tens of seconds. Figure 8 provides a visual demonstration of such bursts, in contrast to the quiet operation in the 'super-stable S2' regime. Both datasets were taken at 2.15 bar under the same voltages as in figure 6 during the 'super-stable S2' phase and its subsequent transition to 'steady state S2'. Each vertical slice provides a color-coded histogram of 2,838 S2 waveforms acquired over ~3 seconds, with ~10 second intervals between consecutive acquisition files. The two lower horizontal strips (in red) are the 59.5 keV peak and the contribution of the low-energy gammas and electrons; the dominant yellow strip is the alpha particle peak. Interestingly, the high-amplitude bursts in (A) have S2 pulse height similar to that in the 'super-stable S2' state.

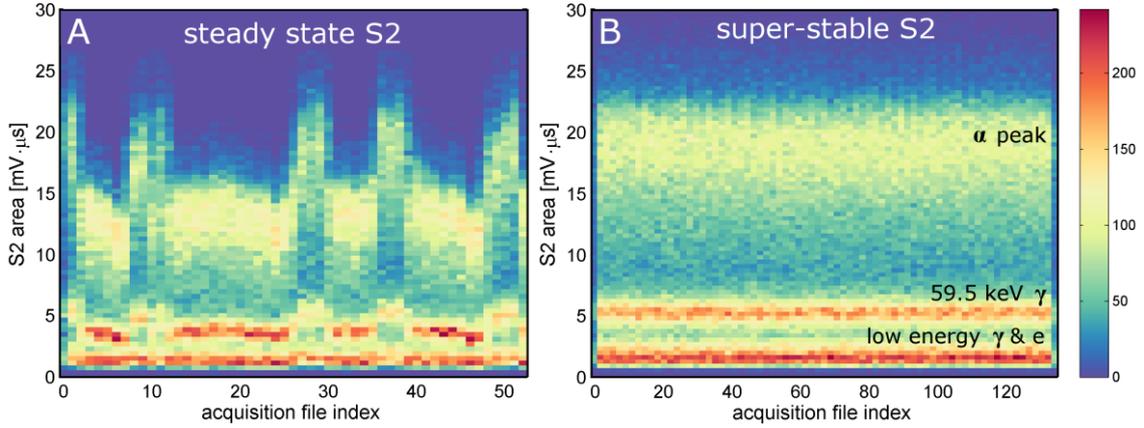

**Figure 8:** Fluctuations in the S2 signal magnitude in steady state (A) and their absence under 'super-stable' conditions (B). Each vertical slice is a color-coded histogram representing an acquisition file comprising 2,838 S2 pulses. Files are shown in consecutive order of acquisition; each acquisition lasted ~3 s, with ~10 s intervals between files. Both datasets (A & B) were recorded under the same voltage and pressure conditions.

Since the 'super-stable' condition in WILiX could not be maintained over more than 1-2 hours and required long periods of cooling down and warming up, a study of the effects of the applied fields and pressure on the S2 signal magnitude and resolution was done in the 'steady state S2' mode; although the width of the S2 distribution was nearly two-fold worse under these conditions, it allowed reaching qualitative conclusions as discussed below. In order to decouple the effects of the voltages and pressure from those of the fluctuations observed in steady state, data analysis was based on selected stable acquisition files (lower-amplitude slices in figure 8A), after excluding files with large-amplitude 'bursts'. The S2-peak centroid and RMS width values were extracted from the stable data following the methodology discussed in section 3.3.1 above.

Data under 'steady state S2' conditions were recorded over a period of ~3 months at both 1.3 bar and 2.1 bar. Figure 9 shows the dependence of the S2 peak centroid on the THGEM voltage at both pressures. All measurements shown in the figure were done with $E_{drift}$ = 0.2-0.3 kV/cm (with the variations resulting from the contribution of the THGEM field) and a PMT voltage of 600 V. Four of the displayed measurements were done with a transfer voltage of −100 V (nominal transfer field $E_{trans}$ = −0.4 kV/cm, pushing the electrons towards the bottom of the THGEM electrode) and one with +250 V ($E_{trans}$ = +1 kV/cm pulling the electrons towards the bottom mesh). At 1.3 bar occasional discharges began to appear at a THGEM voltage of ~2.8 kV; at 2.1 bar they appeared at 3.8 kV. The measurements were thus limited to maximum THGEM voltages of 2.5 kV and 3.5 kV, respectively.

Each point on the curves (figure 9) typically represents ~10,000-30,000 waveforms (after applying the cuts discussed above and discarding files with large fluctuations), taken over a few minutes. The data confirm the previous observations [28] that the electroluminescence process in the THGEM holes is, to first order, linear with the electric field. The curves taken on different dates indicate that over this time span, the system showed robust and reproducible results. A rather surprising conclusion, however, is that the S2 signal magnitude does not appear to depend on the pressure in the vessel (for the two pressures investigated). This is not expected in an electroluminescence process in gas, in which the photon yield scales as $E/\rho$, where $\rho$ is the gas density. A possible explanation is that the curvature of the gas-liquid interface under the THGEM holes is different for the two pressures, affecting the local field inside the gas in a manner that compensates for the gas density effect. Supporting this hypothesis is the dependence of S1 on the system pressure (and THGEM field), discussed in section 3.3.4. Lastly, the direction and strength of the transfer field does not have an observable effect on the magnitude of S2.

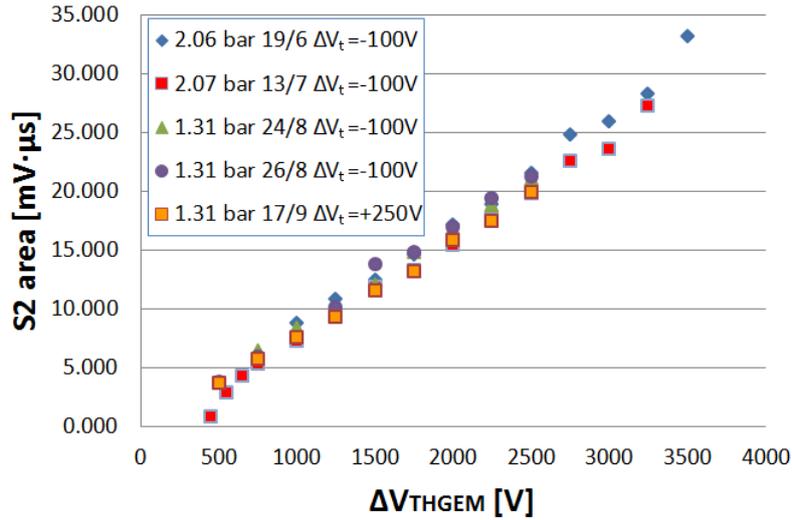

**Figure 9:** The centroid of the alpha-induced S2 vs. the THGEM voltage, at 1.3 bar and 2.1 bar. The measurements were taken over a period of ~3 months (dates are indicated in the legend). In all cases the drift field was 0.2-0.3 kV/cm. The transfer voltage (here denoted $\Delta V_t$) was either −100 V or +250V, with no apparent effect.

Figure 10 shows the combined effect of the drift field and THGEM voltage on the magnitude of the alpha particle-induced S2 signals and their RMS resolution in the 'steady state S2' mode of operation. The data shown were recorded at 2.08 bar, with $E_{drift}$ = 0.2-0.3 kV/cm, 0.5-0.6 kV/cm and 0.9-1.1 kV/cm; the THGEM voltage was scanned from 500 V to 3500 V and a transfer voltage of +250 V (nominal transfer field of 1 kV/cm). A linear dependence on the THGEM voltage is observed in all cases at sufficiently high voltages (figure 10A), with a linear dependence of S2 on the drift field at a fixed THGEM voltage. For a fixed drift field, the RMS S2 resolution (figure 10B) improves with the THGEM voltage until reaching a plateau. The knees in the curves of figure 10A, as well as the beginning of the plateau region in figure 10B, represent the transition from partial to full collection of the ionization electrons into the THGEM holes (as also observed in gas-phase operation of THGEM detectors [36]).

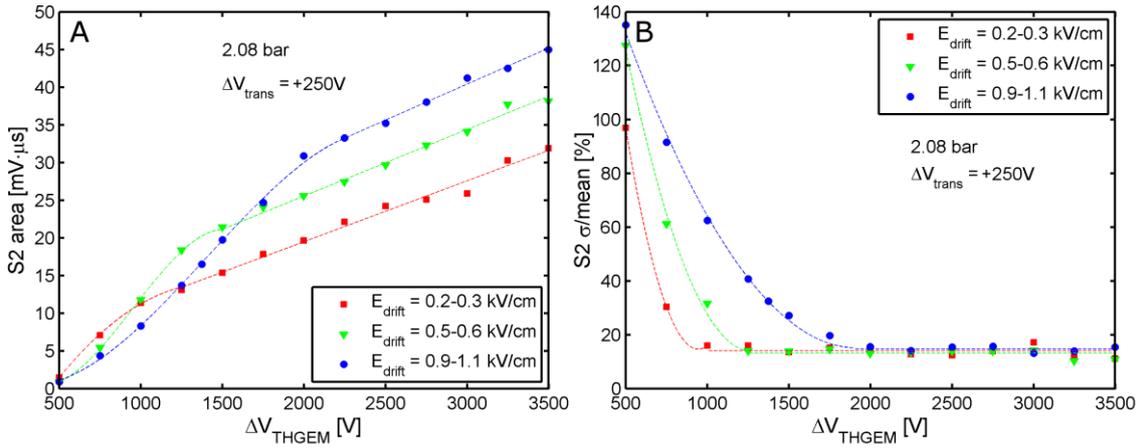

**Figure 10:** The combined effect of the drift field and THGEM voltage on the magnitude (A) and RMS resolution (B) of the alpha-particle S2 signals under 'steady state S2' conditions. The data shown were obtained from stable acquisition files, i.e., excluding the occasional bursts of high-amplitude pulses (figure 8).

3.3.3  *Effect of the transfer field – probing the thickness of the gas layer*

The direction of the transfer field determines whether the trajectory of the electrons drifting out of the holes ends on the THGEM bottom (for a negative transfer voltage), or on the bottom mesh (for a positive transfer voltage). As long as the transfer voltage was below 500 V (nominal transfer field < 2 kV/cm), it appeared to have little effect on the magnitude of the S2 signals and the width of their distribution, regardless of the direction of the transfer field. However, for large positive transfer voltages (> +1000 V, pulling the electrons towards the bottom mesh), we observed a clear effect on the S2 pulse shape, as shown in figure 11. The data comprise a series of S2 pulse shapes (each averaged over several thousand frames), with their amplitude normalized to 1, for positive transfer voltages ranging from 1000 to 2000 V (nominal transfer fields of 4-8 kV/cm). The measurement was done under 'steady state S2' conditions at 2.16 bar with a THGEM voltage of 1000 V and a drift field of 0.2 kV/cm. As indicated in the figure, increasing the transfer voltage from 250 V to 750 V has nearly no effect on the S2 pulse shape. However, for $\Delta V_{trans}$>1000 V the pulse acquires a knee extending ~0.5 μs after the main pulse. Raising the transfer voltage shortens the knee and increases its amplitude until it merges with the main pulse at $\Delta V_{trans}$=2000 V.

The knee structure may be readily explained by the following argument. The scintillation process continues as long as the electrons drift in a field larger than the threshold for scintillation in gas (3.1 kV/cm at 2.16 bar [3]). For low transfer voltages the process ends when the electrons are still inside the bubble/gas layer, and there is no knee. However, for high transfer fields the process continues until the electrons exit the bubble, or until they arrive at the bottom mesh (if the bubble extends below it). Thus, the time from the beginning of the S2 pulse to the knee (figure 11B) is roughly equal to the time required for the electrons to cross the bubble or reach the mesh, and its measurement can be used to estimate the thickness of the gas layer.

Figure 12 shows the measured times from the beginning of the S2 pulse to the knee for $\Delta V_{trans}$=1000 V, 1500 V and 2000 V against the calculated crossing time for gas layers of varying thicknesses (the gas-liquid interfaces were assumed to be flat, with the upper interface at the THGEM bottom; assuming it was half-way into the THGEM hole had negligible effect). The field was calculated using COMSOL and the crossing time was found using the known dependence of the drift velocity in xenon gas on the field [37]. The data agree with a 2.5 mm thick gas layer and cannot be explained by thinner layers or small bubbles. Since the distance between the THGEM and bottom mesh is 2.5 mm, we cannot conclude whether the gas layer extends below the mesh, or is trapped between mesh and THGEM.

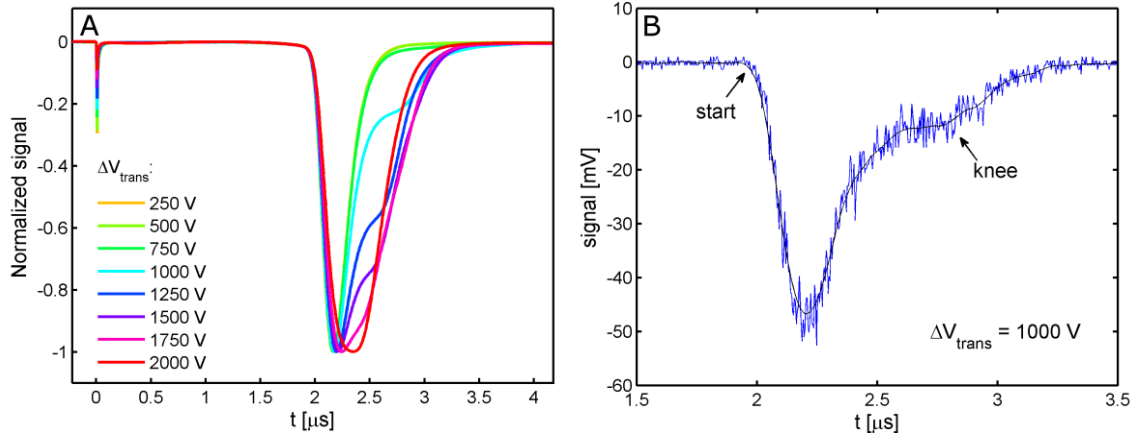

**Figure 11:** (A) S2 signals for varying (positive) transfer voltages, pulling electrons from the bottom of the THGEM hole towards the mesh – presumably through a gas bubble/layer. Each curve represents an average over several thousand waveforms, with the amplitude normalized to unity. For $\Delta V_{trans} \geq 1000$ V a clear knee is observed. (B) A single waveform, showing the 'start' and 'knee' points used for the analysis, taken with $\Delta V_{trans} = 1000$ V. The black curve, obtained by smoothing the data (with a moving average technique), is used as a visual aid for estimating the position of the knee.

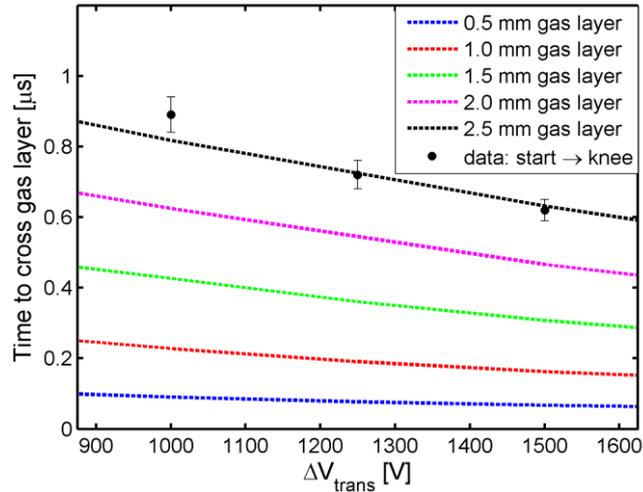

**Figure 12:** The calculated electron crossing times of the gas layer for a varying thicknesses between 0.5 mm and 2.5 mm, compared to the measured time between the S2 pulse rise and knee (figure 11B). The data are consistent with a 2.5 mm thick gas layer, essentially filling the entire gap between the THGEM bottom and the mesh underneath.

*3.3.4 Effect of the THGEM field on S1*

During all measurements − under 'steady state S2', 'super-stable S2' and 'no S2' conditions − waveforms and spectra of S1 scintillation were also recorded. Analysis of the spectra uncovered surprising behavior in the different regimes (figure 13A). The first observation was that under 'no S2' conditions, S1 signals were nearly twice as large as during the 'steady state' or 'super-stable' S2 regimes. A second observation was that while under 'no S2' conditions the THGEM voltage had no effect on S1 (as expected), under 'steady state S2' it did: at 1.3 bar S1 increased with the THGEM voltage, while at 2.1 bar – it decreased; in both cases, the effect was also dependent on the drift field. This peculiar behavior was repeatedly reproduced in many experiments at both pressures.

These observations can be explained by total internal reflection of S1 photons from a curved gas-liquid interface. An S1 photon emitted from the alpha particle track, reaching the THGEM plane at an incidence angle small enough to pass through the THGEM hole towards the PMT (<36.9° for the present THGEM geometry), would be, nearly always, below the critical angle for total internal reflection from a *planar* liquid-gas bubble interface (which, for a refractive index of 1.54-1.69 [3] is between 36.3° and 40.5°); in this case, there would be no difference in S1 with or without bubbles. Only a *curved* liquid gas interface (e.g., as shown in figure 13B) would explain total internal reflection that reduces the number of photons hitting the PMT. The dependence of S1 on the THGEM voltage may be attributed, though further simulations and experimental proof are needed, to the effect of electrostriction [38] (suggested also in [39]). Namely, dipoles in the dielectric liquid induced by the strong electric field inside the holes interact with the field gradient, creating excess pressure in the liquid. The electrostrictive force is of the same order of magnitude as the other relevant forces (pressure inside the bubble and surface tension [33,38]). This would result in a deformation of the bubble shape – possibly in a pressure-dependent manner – leading to a change of the curvature of the total internal reflecting surface and, consequently, to a change in the intensity of light passing through it.

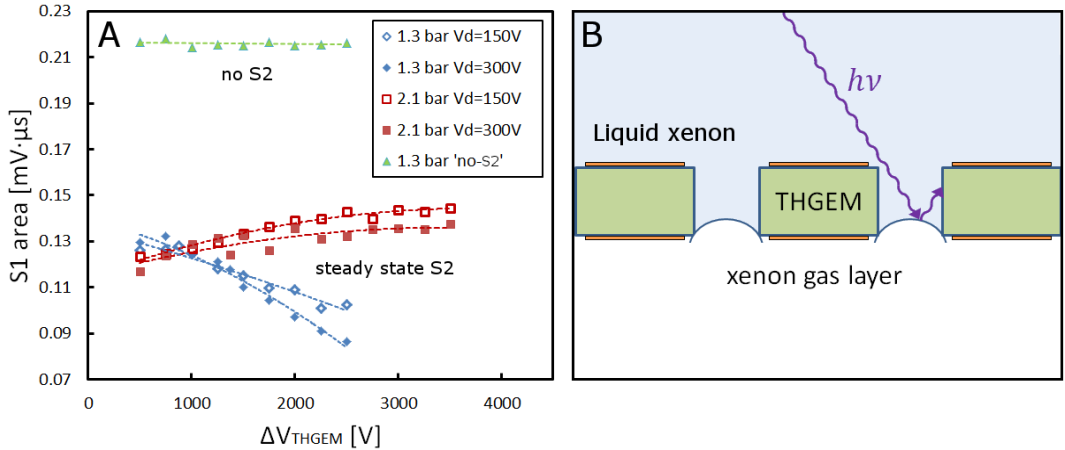

**Figure 13:** (A) Effect of the THGEM voltage on the magnitude of S1 signals under 'no S2' and 'steady state S2' regimes; data shown at 1.3 bar and 2.1 bar for drift voltages of 150 V and 300 V. (B) The two-fold decrease in the S1 magnitude during 'steady state S2' compared to 'no-S2' conditions can be explained by total internal reflection of S1 photons from a curved gas-liquid interface. The change in S1 magnitude for varying THGEM voltages may reflect changes in the curvature of the interface due to electrostriction forces, with different response at low and high pressures.

*3.3.5 Scintillation yield estimates*

Estimating the S2 scintillation yield (i.e., number of photons emitted per drifting electron into 4π) is problematic for two main reasons: (1) we do not know where inside the THGEM hole the scintillation process begins, and (2) the location and shape of the interface between the gas layer and the liquid below are also unknown. We can thus only provide a rough estimate based on some simplifying assumptions. The assumptions we adopt are as follows: (a) all photons are emitted at the bottom of the THGEM hole; (b) the gas-liquid interface coincides with the bottom mesh and is completely flat and specular. Under these assumptions, a simple Monte Carlo calculation that includes Fresnel reflection and refraction at the gas-liquid interface and the mesh transparency (85%) shows that ~27% of the photons emitted into 4π from the hole's bottom reach the PMT.

As noted in section 3.3.1, the measured average number of S1 photons reaching the PMT for a 4.4 MeV alpha particle emitted into the liquid (under 'no-S2' conditions) is approximately 500. According to the data shown in section 3.3.4 (figure 13A), this results in an average S1 area of 0.22 mV·μs. Since the pulse area is proportional to the number of photons reaching the PMT, we can conclude that the response of the PMT used in the experiments (biased at 600 V) is $\sim 4.4 \times 10^{-4}$ mV·μs per photon impinging on its window. Thus, a typical S2 signal of 40 mV·μs (at a THGEM voltage of ~3000 V and drift field of 1 kV/cm) corresponds to $9.1 \times 10^4$ photons hitting the PMT. Since, according to the rough estimate given above, the probability for a photon emitted isotropically from the THGEM bottom to reach the PMT is 27%, a 40 mV·μs signal corresponds to the emission of $3.4 \times 10^5$ photons into 4π. Assuming a charge yield of ~2 electrons/keV from an alpha particle track at 1 kV/cm [2], the number of ionization electrons for a 4.4 MeV alpha is ~8000, and the S2 yield of the THGEM is thus $3.4 \times 10^5 / 8 \times 10^3 \approx 40$ photons/electron.

The above estimate was based on the assumption that all S2 photons are emitted at the bottom of the THGEM hole. If, on the other hand, we had adopted the assumption taken in [28], that *all* the photons are emitted at the center of the hole (but retaining the assumption of a gas layer filling the gap between the THGEM and bottom mesh), the result would be a larger S2 yield of ~140 photons/electron at a THGEM voltage of 3000 V. Thus ~40 photons/electron should be regarded as a conservative estimate, with the actual value possibly larger by up to a factor of ~2. Studies are underway to reduce this uncertainty.

*3.3.6 Additional observations*

In this section we present three auxiliary observations: (1) evidence for modest charge multiplication inside the trapped gas layer; (2) absence of S2 signals when applying a reverse drift field; and (3) (apparent) absence of gas above the THGEM electrode.

To investigate the possibility of charge multiplication, we used a Keithley 610C electrometer to measure the direct current from the THGEM top and bottom surfaces. The measurements, done at 1.30 bar and 2.05 bar under 'steady state S2' conditions, relied on integrating the charge pulses due to the $^{241}$Am alpha and gamma emissions, with typical current values of a few picoamps. The current was first measured from the THGEM top ($I_{top}$) at a drift voltage of 100 V ($E_{drift}$ = 0.3 kV/cm), while applying a reverse voltage (−100 V) across the THGEM itself, to make sure the drifting electrons do not enter its holes. We then changed the scheme to measure the current from the THGEM bottom ($I_{bottom}$), by applying an increasingly larger forward voltage across the THGEM to focus the electrons into the holes, with a reverse transfer voltage of −100 V to repel them to the THGEM bottom. The drift field was kept the same as in the measurement of $I_{top}$. Figure 14 shows the ratio of currents $I_{bottom}/I_{top}$ as a function of the THGEM voltage. For low THGEM voltages, $I_{bottom}/I_{top} < 1$, indicating that a fraction of the electrons is lost to the THGEM top. Starting at $\Delta V_{THGEM} \approx 1$ kV for both pressures, the ratio becomes larger than one, indicating the onset of charge multiplication. The maximum multiplication observed over the stable range of THGEM voltages was ~2 for both pressures. Although a higher charge gain at 1.3 bar at a given voltage can be expected, the curves do not follow a simple *E/ρ* relation, suggesting that additional effects are involved (e.g., a pressure-dependent change in the curvature of the gas-liquid interface, as suggested in section 3.3.4). The observed charge gain is lower, but of the same order of magnitude, as reported in [40] for a single THGEM in xenon gas at low temperatures. This may indicate that the gas-liquid interface is indeed located close to the bottom of the hole such that charge multiplication occurs over a shorter path and in a lower field region of the hole compared to a THGEM operated in gas.

We note that with a charge gain of ~2, an S2 yield of ~50 photons/electron is consistent with that obtained for thin wires in [17], where an S2 yield of ~290 photons/electron was obtained with a charge gain of ~13.

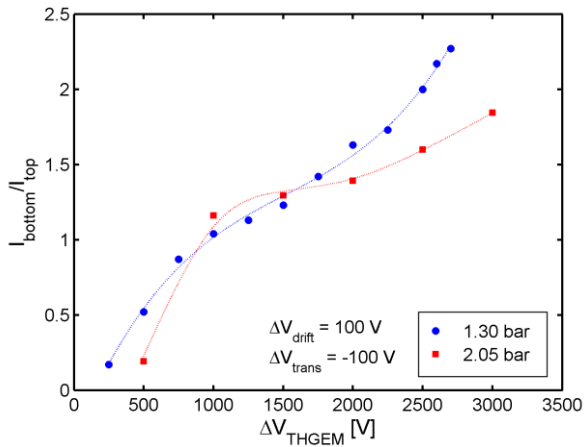

**Figure 14:** Charge gain as a function of the THGEM voltage at 1.30 bar and 2.05 bar under 'steady state S2' conditions. The gain is defined as the ratio between the direct current measured from the THGEM bottom to the current measured from its top at the same drift voltage. The onset of charge multiplication is at $\Delta V_{THGEM} \sim 1$ kV for both pressures.

A second observation concerns the question whether S2 signals appear in the absence of a drift field (or, more precisely, when the drift field is reversed). Such signals may indicate that there is occasional field emission of electrons inside a hole (or close to its upper rim), and would contribute unwanted 'dark S2 pulses.' To test this, we applied a reversed drift voltage of −100 V ($E_{drift}$ = −0.3 kV/cm) and scanned the THGEM voltage from 500 V to 3500 V in 'steady state S2' conditions at 2.07 bar, maintaining a reversed transfer voltage of −100 V. At each voltage step we acquired 14,190 waveforms, with a low trigger level, essentially capturing the entire alpha-induced S1 signal spectrum. The percentage of triggers on S2 signals (identified by their much larger width compared to S1) increased from 0.01% at a THGEM voltage of 1000 V to 0.11% at 3500 V. Since the rate of alpha particle emission into the liquid was ~8 kHz, this corresponds to a rate of ~1-10 Hz of S2 signals above threshold. Most of these S2 signals were either not accompanied by an S1 signal, or started simultaneously with S1; in less than ~10% of the triggers, a clear S1 was seen before S2, with a typical time difference of ~300-500 ns. At this stage we cannot conclude whether these observed rare S2 signals are in fact dark pulses, as they may have been caused by either cosmic muons passing through the THGEM holes or radioactive emissions (for example, from the THGEM itself) depositing charge inside the hole. The increase in the rate of these S2 signals with the THGEM voltage can be attributed to the increase in the size of the drift 'funnel' above the holes.

The last observation concerns the question of the presence of gas above the THGEM (i.e., in the drift region between the THGEM and source). To investigate this, we applied a reversed voltage across the THGEM (−100 V) and gradually increased the drift voltage from 2000 to 3600 V (corresponding to a drift field of 5.9 – 10.6 kV/cm, which is high enough to ensure practically 100% efficiency of electron extraction from the liquid to gas). During the experiment (which lasted ~15 minutes), there were two cases in which, for a few seconds, small S2-like signals could be seen and gradually faded away; we interpret those as bubbles which were momentarily trapped below the source and gradually collapsed or moved away. Other than these two isolated cases, there was no sign of S2 signals. This leads to the conclusion that under steady-state conditions the gas layer remains trapped below the THGEM, with rare release of bubbles into the drift gap (not necessarily through the THGEM holes).

## 4 Summary and discussion

This work was initiated as part of our effort to investigate the feasibility of the liquid-hole multiplier (LHM) concept that was conceived as a possible means for detecting ionization and scintillation signals in large-scale single-phase noble-liquid TPCs [18]. Our initial results with a THGEM immersed in liquid xenon [28], which showed large alpha particle-induced S2 ionization signals scaling linearly with the applied voltage, were highly encouraging. However, the threshold field at which the electroluminescence process first appeared (a few kV/cm) was surprisingly low – roughly two orders of magnitude below the one reported for thin wires in LXe. As detailed above, further investigation of this result leads us to conclude that the S2 light was, in fact, produced in a gas layer/bubble trapped below the THGEM electrode, rather than in the liquid itself.

Although final proof of the bubble hypothesis would require their direct observation, the indirect evidence gathered so far is compelling. In our view, the most convincing result is the disappearance and reappearance of the S2 signals following rapid pressure changes in the system. This is further supported by the two-fold increase in the S1 scintillation signal when switching from 'steady state S2' to 'no S2' conditions, which we explained by total internal reflection of S1 photons at the curved gas-liquid interface inside the THGEM holes. A third supporting evidence is the knee observed when applying a strong transfer field between the THGEM and bottom mesh. And finally, the few-kV/cm threshold for scintillation observed in [28] and reproduced in this work is exactly as would be expected for scintillation in xenon vapor.

The natural expectation concerning a bubble-based phenomenon is that the process would be inherently unstable. Indeed, under 'steady state S2' conditions considerable fluctuations are observed and the S2 RMS resolution, without discarding the fluctuations, is $\sigma/E$~20-25%. While one may conceive applications where this would be sufficient, it presents a step back compared to the performance of present dual-phase LXe TPCs. However, the existence of the 'super-stable S2' condition (showing an S2 resolution of ~11%, similar to that of the XENON100 dark matter experiment), leaves room for optimism; it demonstrates that such conditions are achievable. The challenge, of course, will be to control the bubble-formation process in order to stay in the observed 'super-stable S2' regime in steady state.

The study conducted with respect to the roles played by the electric fields in the THGEM and in the drift gap shows that full collection of the drifting ionization electrons into the THGEM holes can be readily achieved, with the THGEM operating in a fully stable regime. Even though the S2 yield deduced in the present work, $\gtrsim 40$ photons (emitted into $4\pi$) per drifting electron, is a few times lower than that reported for thin wires, it is still quite high and should allow for efficient detection of events comprising a small number of ionization electrons. It still remains to be seen whether *single* electrons can also be detected efficiently using this method. We note that unlike the case of thin wires, where the recorded light signal is critically dependent on the trajectory of the drifting electrons relative to the wire (because of shadowing by the wire itself), there is essentially no such dependence in the case of a bubble-assisted LHM.

Although we do not yet have a complete picture of the structure of the trapped gas pocket (bubbles or layer), or where the bubbles are nucleated, the data appear consistent with a full gas layer between the THGEM electrode and the bottom mesh; whether the layer extends below the mesh, or is trapped between the THGEM and mesh is still unclear. If the latter is true, it may open up possibilities for controlled trapping of xenon gas between perforated electrodes (THGEMs, GEMs, meshes, etc.), perhaps not only in a horizontal orientation. Moreover, as noted in section 3.3.4, the intriguing effect of the THGEM voltage on the S1 signal magnitude may suggest that electrostriction forces are in play. This may provide an additional handle for manipulating the trapped gas layer and perhaps help in devising ways of keeping it in the 'super-stable S2' mode.

If a practical scheme to maintain the 'super-stable' condition can be devised, one can conceive various designs for large-scale single-phase noble-liquid TPCs, where S2 signals are produced in locally confined gas pockets. One such example is shown in figure 15A, which depicts a liquid-only TPC design where ionization electrons drift down towards an LHM array, with a controllably produced gas layer underneath (for example, by a set of heating wires). The LHM in this case may incorporate a CsI photocathode on its top surface (as first suggested in [23] and shown in figure 15B) to allow for the detection of S1 photons in addition to ionization electrons. Light readout of the LHM signals can be performed either by PMTs or, preferably, by pixilated photosensors (e.g., GPMs [41-43] or SiPMs). We note that for the bubble-assisted LHM idea to be applicable to large volume TPCs, the bubbles must be effectively trapped to prevent them from reaching high voltage regions where they may induce electrical breakdowns.

The work presented here was performed in liquid xenon; nevertheless, its results may apply to liquid argon as well. Indeed, the fairly low-field scintillation threshold reported for a THGEM [30] and GEM [31] immersed in LAr may have also resulted from accidental formation of argon bubbles. Moreover, in a work on electron avalanche multiplication in LAr on a sharp needle tip [44], the authors have identified a pressure-dependent contribution to the signal and raised the suspicion of bubble formation at the tip. A possible application of bubble-assisted LHMs may be in LAr TPCs for neutrino physics experiments; LHMs at the bottom of the TPC with light readout by pixilated photosensors (as suggested in [30] and similar to the scheme of figure 15A) would allow for highly accurate track reconstruction, with additional information on the track charge density and the ability to operate in magnetic fields.

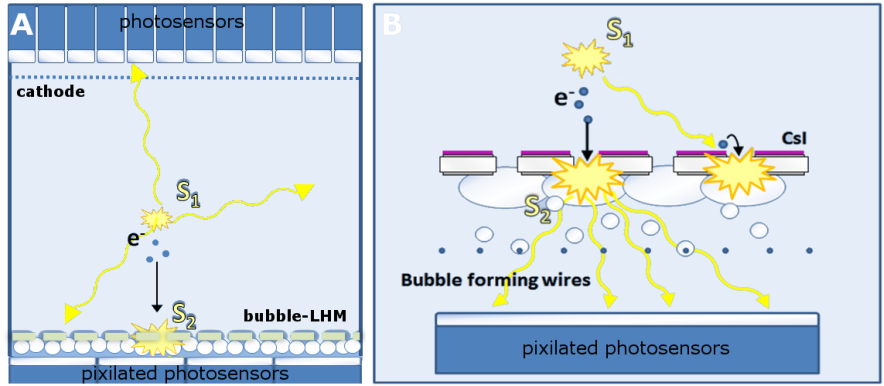

**Figure 15:** (A) Schematic design of a liquid-only single-phase TPC with bubble-assisted LHMs at the bottom. S2 signals are created by electrons drifting down towards the LHM. Accurate S2-based position reconstruction is permitted by pixilated photosensors below the LHM. (B) A single LHM element, coated with CsI to allow for the detection of S1 photons in addition to ionization electrons. Controlled bubble formation can be done by ohmic heating of resistive wires below the electrode.

## Acknowledgements

This work was partly supported by the Minerva Foundation with funding from the German Ministry for Education and Research (Grant No. 710827), the Israel Science Foundation (Grant No. 477/10) and by FCT and FEDER under the COMPETE program, through project CERN/FP/123614/2011. A. Breskin is the W.P. Reuther Professor of Research in The Peaceful Use of Atomic Energy. We thank Mr. B. Pasmantirer and Mr. S. Assayag and members of their respective Design Office and Mechanical Workshop at the Weizmann Institute, for their invaluable assistance in the design and manufacture of the experimental setup. V. Chepel expresses his thanks to the Weizmann Institute's Physics Faculty for hospitality and financial support. The research has been carried out within the DARWIN Consortium for future dark matter experiments.

## References

[1]   B. A. Dolgoshein, V. N. Lebedenko and B. U. Rodionov, *New Method of Registration of Ionizing-particle Tracks in Condensed Matter*, JETP Letters **11**(1970) 351.

[2]   E. Aprile and T. Doke, *Liquid xenon detectors for particle physics and astrophysics*, Rev. Mod. Phys. **82**(2010) 2053.

[3]   V. Chepel and H. Araujo, *Liquid noble gas detectors for low energy particle physics*, 2013 *JINST* **8** R04001.

[4]   E. Aprile, et al., *The XENON100 dark matter experiment*, Astropart. Phys. **35**(2012) 573.

[5]   E. Aprile, et al., *Dark Matter Results from 225 Live Days of XENON100 Data*, Phys. Rev. Lett. **109**(2012) 181301.

[6]   D. S. Akerib, et al., *The Large Underground Xenon (LUX) experiment*, Nucl. Instrum. Meth. A **704**(2013) 111.

[7]   D. S. Akerib, et al., *First Results from the LUX Dark Matter Experiment at the Sanford Underground Research Facility*, Phys. Rev. Lett. **112**(2014) 091303.

[8]   X. G. Cao, et al., *PandaX: a liquid xenon dark matter experiment at CJPL*, Sci. China Phys. Mech. **57**(2014) 1476.

[9]  M. J. Xiao, et al., *First dark matter search results from the PandaX-I experiment, Sci. China Phys. Mech.* **57**(2014) 2024.

[10] T. Alexander, et al., *DarkSide search for dark matter,* 2013 *JINST* **8** C11021.

[11] E. Aprile for the XENON1T collaboration, *The XENON1T Dark Matter Search Experiment*, *Springer Proc. Phys.* **148** (2013) 93. arXiv:1206.6288 [astro-ph.IM]

[12] A. Badertscher, et al., *ArDM: first results from underground commissioning,* 2013 *JINST* **8** C09005.

[13] D. C. Malling, et al., *After LUX: The LZ Program.* arXiv:1110.0103 [astro-ph.IM]

[14] L. Baudis for the DARWIN Consortium, *DARWIN dark matter WIMP search with noble liquids, J. Phys. Conf. Ser.* **375**(2012) 2028.

[15] P. Majewski., *R&D for future ZEPLIN: a multi-ton LXe DM detector*. in *CRYODET – Cryogenic Liquid Detectors for Future Particle Physics*. 2006. LNGS, Italy.

[16] K. L. Giboni, X. Ji, H. Lin and T. Ye, *On Dark Matter detector concepts with large-area cryogenic Gaseous Photo Multipliers,* 2014 *JINST* **9** C02021.

[17] E. Aprile, H. Contreras, L. W. Goetzke, A. J. Melgarejo Fernandez, M. Messina, J. Naganoma, G. Plante, A. Rizzo, P. Shagin and R. Wall, *Measurements of proportional scintillation and electron multiplication in liquid xenon using thin wires,* 2014 *JINST* **9** P11012.

[18] T. Ye, K. L. Giboni and X. Ji, *Initial evaluation of proportional scintillation in liquid Xenon for direct dark matter detection,* 2014 *JINST* **9** P12007.

[19] S. E. Derenzo, T. S. Mast, H. Zaklad and R. A. Muller, *Electron avalanche in liquid xenon, Phys. Rev. A* **9**(1974) 2582.

[20] M. Miyajima, K. Masuda, A. Hitachi, T. Doke, T. Takahashi, S. Konno, T. Hamada, S. Kubota, A. Nakamoto, E. Shibamura, *Proportional counter filled with highly purified liquid xenon, Nucl. Instrum. Meth.* 134(1976) 403.

[21] K. Masuda, S. Takasu, T. Doke, T. Takahashi, A. Nakamoto, S. Kubota, E. Shibamura, *Liquid Xenon Proportional Scintillation-Counter, Nucl. Instrum. Meth.* **160**(1979) 247.

[22] T. Doke, *Recent developments of liquid xenon detectors, Nucl. Instrum. Meth.* **196**(1982) 87.

[23] A. Breskin, *Liquid Hole-Multipliers: A potential concept for large single-phase noble-liquid TPCs of rare events, J. Phys. Conf. Ser.* **460**(2013) 012020.

[24] F. Sauli, *GEM: A new concept for electron amplification in gas detectors, Nucl. Instrum. Meth. A* **386**(1997) 531.

[25] R. Chechik, A. Breskin, C. Shalem and D. Mormann, *Thick GEM-like hole multipliers: properties and possible applications, Nucl. Instrum. Meth. A* **535**(2004) 303.

[26] Breskin, R. Alon, M. Cortesi, R. Chechik, J. Miyamoto, V. Dangendorf, J. Maia and J. M. F. Dos Santos,, *A concise review on THGEM detectors, Nucl. Instrum. Meth. A* **598**(2009) 107.

[27] R. Chechik and A. Breskin, *Advances in gaseous photomultipliers, Nucl. Instrum. Meth. A* **595**(2008) 116.

[28] L. Arazi, A. E. C. Coimbra, R. Itay, H. Landsman, L. Levinson, B. Pasmantirer, M. L. Rappaport, D. Vartsky and A. Breskin, *First observation of liquid-xenon proportional electroluminescence in THGEM holes,* 2013 *JINST* **8** C12004.

[29] A. Lansiart, A. Seigneur, J. L. Moretti and J. P. Morucci, *Development Research on a Highly Luminous Condensed Xenon Scintillator, Nucl. Instrum. Meth.* **135**(1976) 47.

[30] P. K. Lightfoot, G. J. Barker, K. Mavrokoridis, Y. A. Ramachers and N. J. C. Spooner, *Optical readout tracking detector concept using secondary scintillation from liquid argon generated by a thick gas electron multiplier,* 2009 *JINST* **4** P04002.

[31] A. Buzulutskov, *Advances in Cryogenic Avalanche Detectors,* 2012 *JINST* **7** C02025 arXiv:1112.6153 [physics.ins-det].

[32] E. M. Gushchin, A. A. Kruglov and I. M. Obodovski, *Electron dynamics in condensed argon and xenon.* JETP, 1982. **55**(4) 650.

[33] NIST, *Thermophysical properties of fluid systems, http://webbook.nist.gov/chemistry/fluid/*

[34] E. Aprile, R. Mukherjee and M. Suzuki, *Ionization of liquid xenon by $^{241}$Am and $^{210}$Po alpha particles, Nucl. Instrum. Meth. A:* **307**(1991) 119.

[35] T. Ya. Voronova, M. A. Kirsanov, A. A. Kruglov, I. M. Obodovski, S. G. Pokachalov, V. A. Shilov, and E. B. Khristich, *Ionization yield from electron tracks in liquid xenon, Sov. Phys. Tech. Phys.* **34**(1989) 825.

[36] M. Cortesi, V. Peskov, G. Bartesaghi, J. Miyamoto, S. Cohen, R. Chechik, J. M. Maia, J. M. F. dos Santos, G. Gambarini ,V. Dangendorf and A. Breskin, *THGEM operation in Ne and Ne/CH$_4$*, 2009 *JINST* **4** P08001.

[37] H. L. Brooks, M. C. Cornell, J. Fletcher, I. M. Littlewood and K. J. Nygaard, *Electron-Drift Velocities in Xenon, J. Phys. D Appl. Phys.* **15**(1982) L51.

[38] W. F. Schmidt, *Liquid State Electronics of Insulating Liquids*. 1997, Boca Raton, Florida: CRC Press LLC.

[39] B. A. Dolgoshein, A. A. Kruglov, V. N. Lebedenko, V. P. Miroshnichenko and B. U. Rodionov, *Electronic method of particle detection in double phase liquid-gas systems, Physics of Elementary Particles and Atomic Nucleus* **4**(1973) 20.

[40] A. Bondar, A. Buzulutskov, A. Grebenuk, E. Shemyakina, A. Sokolov, D. Akimov, I. Alexandrov and A. Breskin, *On the low-temperature performances of THGEM and THGEM/G-APD multipliers in gaseous and two-phase Xe,* 2011 *JINST* **6** P07008.

[41] L. Arazi, A. E. C. Coimbra, E. Erdal, I. Israelashvili, M. L. Rappaport, S. Shchemelinin, D. Vartsky, J. M. F. dos Santos and A. Breskin, *Cryogenic gaseous photomultipliers and liquid hole-multipliers: advances in THGEM-based sensors for future noble-liquid TPCs*. Submitted to the proceedings of the *Seventh Symposium on Large TPCs for Low Energy Rare Event Detection*. 2015. https://indico.cern.ch/event/340656/contribution/46/material/slides/1.pdf

[42] A. Breskin, V. Peskov, M. Cortesi, R. Budnik, R. Chechik, S. Duval, D. Thers, A. E. C. Coimbra, J. M. F. dos Santos, J. A. M. Lopes, C. D. R. Azevedo and J. F. C. A. Veloso, *CsI-THGEM gaseous photomultipliers for RICH and noble-liquid detectors, Nucl. Instrum. Meth. A* **639**(2011) 117.

[43] S. Duval , L. Arazi, A. Breskin, R. Budnik, W. T. Chen, H. Carduner, A. E. C. Coimbra, M. Cortesi, R. Kaner, J. P. Cussonneau, J. Donnard, J. Lamblin, O. Lemaire, P. Le Ray, J. A. M. Lopes, A. F. M. Hadi, E. Morteau, T. Oger, J. M. F. dos Santos, L. S. Lavina, J. S. Stutzmann, D. Thers, *Hybrid multi micropattern gaseous photomultiplier for detection of liquid-xenon scintillation, Nucl. Instrum. Meth. A* **695**(2012) 163.

[44] J. G. Kim, S. M. Dardin, R. W. Kadel, J. A. Kadyk, V. Peskov and W. A. Wenzel, *Electron avalanches in liquid argon mixtures, Nucl. Instrum. Meth. A* **534**(2004) 376.